\documentstyle[epsfig,preprint,aps]{revtex}
\begin{document}
\tighten
\title{Effect of bound nucleon internal structure change
on nuclear structure functions}

\author{K. Tsushima$^{a,b,c}$, K. Saito$^{d}$, 
F.M. Steffens$^{a,b}$\vspace{3ex}}
\address{$^a$Instituto de F\'{\i}sica Te\'orica - UNESP, Rua Pamplona 145,
01405-900, S\~ao Paulo, SP, Brazil}
\address{$^b$Mackenzie Presbiteriana University - FCBEE,
Rua da Consola\c{c}\~ao 930,
01302-907, S\~ao Paulo, SP, Brazil}
\address{$^c$National Center for Theoretical Sciences at Taipei, 
Taipei 10617, Taiwan}
\address{$^d$Department of Physics, Faculty of Science and Technology,
Tokyo University of Science, Noda 278-8510, Japan}

\maketitle


\begin{abstract}
Effect of bound nucleon internal structure change on nuclear
structure functions is investigated based on local quark-hadron
duality. The bound nucleon structure functions calculated for
charged-lepton and (anti)neutrino scattering are all enhanced in
symmetric nuclear matter at large Bjorken-$x$ ($x \agt 0.85$)
relative to those in a free nucleon. This implies that a part of
the enhancement observed in the nuclear structure function $F_2$
(in the resonance region) at large Bjorken-$x$ (the EMC effect) is
due to the effect of the bound nucleon internal structure change.
However, the $x$ dependence for the charged-lepton and
(anti)neutrino scattering is different. The former [latter] is
enhanced [quenched] in the region $0.8 \alt x \alt 0.9$ [$0.7 \alt
x \alt 0.85$] due to the difference of the contribution from axial
vector form factor. Because of these differences charge symmetry
breaking in parton distributions will be enhanced in nuclei.

\vspace{1em}
\end{abstract}
%

\section{Introduction}

Recent precision data on the $F_2$ structure function in the
resonance region from Jefferson Lab~\cite{JLab}, as well as the
spin asymmetry ($A_1$) data in the resonance region from
HERMES~\cite{A1}, have significantly renewed the
interest~\cite{Ent,Wally,Close,WKT,Arrington,Wally2} for local
quark-hadron duality (local duality)~\cite{Bloom}. Before the
advent of QCD, it was empirically observed by Bloom and
Gilman~\cite{Bloom} that the nucleon structure function in the
resonance region, $F_2^{res}$, was approximately equal to the
scaling structure function, $F_2^{sca}$, measured in the deep
inelastic region. The relation between the two structure functions
was found to be in terms of a finite-energy sum rule.
Specifically, it was observed that $F_2^{res}$ oscillates around
the scaling curve, and its average is equivalent to $F_2^{sca}$
within a deviation of $\sim 10$ \%. The QCD justification for the
Bloom and Gilman observation, named local duality, was made by De
R\'{u}jula, Georgi, and Politzer~\cite{Rujula}. More recent
theoretical studies of local duality and QCD also have been
made~\cite{Carlson,Ji}.

On the other hand, the European Muon Collaboration (EMC)
effect~\cite{EMC,EMC2} shows that the parton distributions in
nuclei (bound nucleon) are different from those in a free nucleon.
Basically, there happens a depletion of the structure function at
intermediate and larger Bjorken-$x$ ($0.5 \alt x \alt 0.85$), 
while a steep rise at large Bjorken-$x$.
It was concluded by Miller and Smith~\cite{Miller} that the
depletion of the deep inelastic nuclear structure functions
observed in the valence quark regime, is due to some effect beyond
the conventional nucleon-meson treatment of nuclear
physics~\cite{ST}.
We note that for the region of $x > 1$ 
short-range correlations as well as the nuclear binding effects become
important~\cite{SaitoF}.

Recent experimental data from
Jefferson Lab~\cite{Arrington} for ratios of the nuclear to
deuterium cross sections in the resonance region, revealed a
similar effect as that observed in the EMC effect. Because the
resonance structure functions average better in nuclei due to
Fermi motion and nuclear many-body effects~\cite{Arrington}, we
expect that local duality is realized better in nuclei, although a
proper treatment of resonances in nuclei must be investigated.
Thus, local duality may potentially be one of the other new
methods to study the nuclear structure functions at large
Bjorken-$x$.

Complementary, the measurements of electromagnetic form factors of
bound protons in polarized ($\vec e, e' \vec p$) scattering on
$^4$He at MAMI and Jefferson Lab~\cite{He4}, concluded that the
ratio of the electric ($G^p_E$) to magnetic ($G^p_M$) Sachs proton
form factors differs by $\sim 10$ \% in $^4$He from that in $^1$H.
This strongly suggests that the internal structure of a bound
proton (nucleon) is modified in nuclei. In their analyses,
conventional models employing free proton form factors,
phenomenological optical potentials, and bound state wave
functions, as well as relativistic corrections, meson exchange
currents, isobar contributions and final state
interactions~\cite{He4,KELLY}, all fail to account for the
observed effect in $^4$He~\cite{He4}. Indeed, a better agreement
with the data was obtained only when, in addition to these
standard nuclear-structure corrections, a correction due to the
internal structure change of a bound proton in $^4$He was taken
into account~\cite{He4,QMC_EMff}. 
In Ref.~\cite{WKT}, one of such attempts was made to
extract the bound nucleon (nuclear) structure function $F_2$ at
large Bjorken-$x$.

In this Letter, we extend the study of Ref.~\cite{WKT} to
calculate also other bound nucleon structure functions for both
charged-lepton and (anti)neutrino scattering based on local
duality. In particular, we focus on the effect of the bound
nucleon internal structure change on the nuclear structure
functions. However, one of the important, 
conventional nuclear effects, 
Fermi motion, will be included in a simple Fermi gas model. 
We convolute the structure functions with 
the nucleon momentum distribution obtained in a Fermi gas model, 
for {\it both} the structure functions extracted using 
the free and bound nucleon form factors, and then calculate ratios.
(To draw more solid conclusions, we need to perform an elaborated
study including the nuclear effects. We plan to study the effects
of binding etc. in the framework of local quark-hadron
duality~\cite{Saito}.) In addition, we consider charge symmetry
breaking in parton distributions in nuclei using the structure
functions calculated. For this purpose we use the bound nucleon
form factors calculated~\cite{QMC_EMff,neuA} in the quark-meson
coupling (QMC) model~\cite{QMC}, which has been successfully
applied to many problems of nuclear and hadronic
physics~\cite{QMC_app}.

Since, in the present study, we want to see the effect of the
bound nucleon internal structure change included entirely in the
bound nucleon form factors and the pion threshold in a nuclear
medium, the elastic contribution to the bound nucleon structure
functions for charged-lepton scattering may be given
by~\cite{Wally,Carlson}:
\begin{eqnarray}
F_1^{BN*} &=& \frac{1}{2} (G^*_M)^2
\delta(x-1),
\label{F_1BN} \\
F_2^{BN*} &=&
\frac{1}{1+\tau} \left[(G^*_E)^2 + \tau (G^*_M)^2\right]
\delta(x-1),
\label{F_2BN} \\
g_1^{BN*} &=&
\frac{1}{2(1+\tau)} G^*_M \left(G^*_E + \tau G^*_M\right)
\delta(x-1),
\label{g_1BN} \\
g_2^{BN*} &=&
\frac{\tau}{2(1+\tau)} G^*_M \left(G^*_E - G^*_M\right)
\delta(x-1),
\label{g_2BN}
\end{eqnarray}
while those for (anti)neutrino scattering for an isoscalar nucleon,
$N \equiv \frac{1}{2}(p + n)$, may be given by~\cite{FK}:
\begin{eqnarray}
F_1^{WNBN*}
&=& \frac{1}{4} \left[(G^{V*}_M)^2 + (1 + 1/\tau)(G^*_A)^2 \right]
\delta(x-1),
\label{F_1WNBN} \\
F_2^{WNBN*}
&=& \frac{1}{2} \left[\frac{(G^{V*}_E)^2 + \tau (G^{V*}_M)^2}{1 + \tau}
+ (G^*_A)^2 \right]
\delta(x-1),
\label{F_2WNBN} \\
F_3^{WNBN*}
&=& G^{V*}_MG^*_A
\delta(x-1),
\label{F_3WNBN}
\end{eqnarray}
where $\tau = Q^2/4M^2$ ($M$, the free nucleon mass) and $x$ the
Bjorken variable. $G^*_E$ $[G^*_M]$ is the bound nucleon electric
[magnetic] Sachs form factor~\cite{QMC_EMff}, $G^{V*}_{E,M} =
G^{p*}_{E,M} - G^{n*}_{E,M}$ the corresponding isovector
electromagnetic form factors, and $G^*_A$ is the (isovector) axial
vector form factor~\cite{neuA,FK}.

Using the Nachtmann variable, $\xi = 2x/(1+\sqrt{1+x^2/\tau})$,
local duality equates the scaling bound nucleon structure function
$F_2^*$ and the contribution from $F_2^{BN*}$ of
Eq.~(\ref{F_2BN}):
\begin{equation}
\int_{\xi^*_{th}}^1 F_2^*(\xi) d\xi
=
\int_{\xi^*_{th}}^1 F_2^{BN*}(\xi,Q^2) d\xi,
\label{LQHD}
\end{equation}
where $\xi^*_{th}$ is the value at the pion threshold in a nuclear
medium given below. In this study we consider symmetric nuclear
matter, and we can assume that the pion mass in-medium ($m^*_\pi$) is
nearly equal to that in free space ($m_\pi$), and $\xi^*_{th}$ will
be given by~\cite{WKT}:
\begin{equation}
\xi^*_{th} = \xi (x^*_{th}), \label{xi_th_medium}
\end{equation}
with
\begin{equation}
x^*_{th} =
x_{th} \frac{m_\pi (2M + m_\pi) + Q^2}{m_\pi [2(M^* + 3 V^q_\omega) + m_\pi]
+ Q^2},
\qquad
x_{th} = \frac{Q^2}{m_\pi (2M + m_\pi) + Q^2},
\label{x_th_medium}
\end{equation}
where $x^*_{th}$ [$x_{th}$] is the Bjorken-$x$ at the
pion threshold in medium [free space], and
$M^*$ and $3 V^q_\omega$~\cite{QMC} are respectively
the effective mass and the vector potential of the bound nucleon.
Inserting Eq.~(\ref{F_2BN}) into the r.h.s. of Eq.~(\ref{LQHD}),
we get~\cite{Wally,Rujula}:
\begin{equation}
\int_{\xi^*_{th}}^1 F_2^*(\xi) d\xi
=
\frac{\xi_0^2}{4-2\xi_0}
\left[ \frac{(G^*_E)^2 + \tau (G^*_M)^2}{1 + \tau} \right],
\label{F_2_LQHD}
\end{equation}
where $\xi_0 = \xi (x = 1)$.
The derivative in terms of $\xi^*_{th}$ in both sides of Eq.~(\ref{F_2_LQHD})
with $\xi_0$ fixed gives~\cite{Wally,FK}:
\begin{equation}
F_2^*(\xi^*_{th}) \equiv F_2^*(x^*_{th})
=
-2\beta^* \left[\frac{(G^*_M)^2 - (G^*_E)^2}{4M^2(1 + \tau )^2}
+ \frac{1}{1 + \tau}
\left(\frac{d(G^*_E)^2}{dQ^2} + \tau\frac{d(G^*_M)^2}{dQ^2}\right)\right],
\label{F_2_x_th}
\end{equation}
where $\beta^* =
(Q^4/M^2)(\xi^2_0/\xi^{*3}_{th})[(2-\xi^*_{th}/x^*_{th})/(4-2\xi_0)]$.
Our $\beta^*$ is different from that in Ref.~\cite{Wally} by an extra
factor $1/x_{th}$ in the limit of zero baryon density. In
addition, we have an overall minus sign in the expression of
Eq.~(\ref{F_2_x_th}). However, because only ratios are calculated
in Ref.~\cite{Wally} the conclusions in there are unaffected.
(These as well as the expression for $g_2$ below, 
have been corrected by the author~\cite{Wally}.) 
Similarly, we get expressions for other bound nucleon structure
functions at $x = x^*_{th}$:
\begin{eqnarray}
F_1^*(x^*_{th}) &=& -\beta^*\frac{d(G^*_M)^2}{dQ^2},
\label{F_1_x_th}\\
g_1^*(x^*_{th})
&=&
-\beta^* \left[\frac{G^*_M(G^*_M - G^*_E)}{4M^2(1 + \tau )^2}
+ \frac{1}{1 + \tau}
\left(\frac{d(G^*_E G^*_M)}{dQ^2} + \tau\frac{d(G^*_M)^2}{dQ^2}\right)\right],
\label{g_1_x_th}\\
g_2^*(x^*_{th})
&=&
-\beta^* \left[\frac{G^*_M(G^*_E - G^*_M)}{4M^2(1 + \tau )^2}
+ \frac{\tau}{1 + \tau}
\left(\frac{d(G^*_E G^*_M)}{dQ^2} - \frac{d(G^*_M)^2}{dQ^2}\right)\right],
\label{g_2_x_th}\\
F_1^{WN*}(x^*_{th})
&=&
-\frac{\beta^*}{2} \left[\frac{-(G^*_A)^2}{4M^2\tau^2}
+ \frac{d(G^{V*}_M)^2}{dQ^2}
+ \frac{1 + \tau}{\tau}\frac{d(G^*_A)^2}{dQ^2} \right],
\label{F_1WN_x_th}\\
F_2^{WN*}(x^*_{th})
&=&
-\beta^*\left[\frac{(G^{V*}_M)^2 - (G^{V*}_E)^2}{4M^2(1 + \tau )^2}
+ \frac{1}{1 + \tau}
\left(\frac{d(G^{V*}_E)^2}{dQ^2} + \tau\frac{d(G^{V*}_M)^2}{dQ^2}\right)
+ \frac{d(G^*_A)^2}{dQ^2} \right],
\label{F_2WN_x_th}\\
F_3^{WN*}(x^*_{th})
&=&
-\beta^* \frac{(2G^{V*}_MG^*_A)}{dQ^2}.
\label{F_3WN_x_th}
\end{eqnarray}
%

\begin{figure}
\begin{center}
\vspace*{1cm}
\epsfig{file=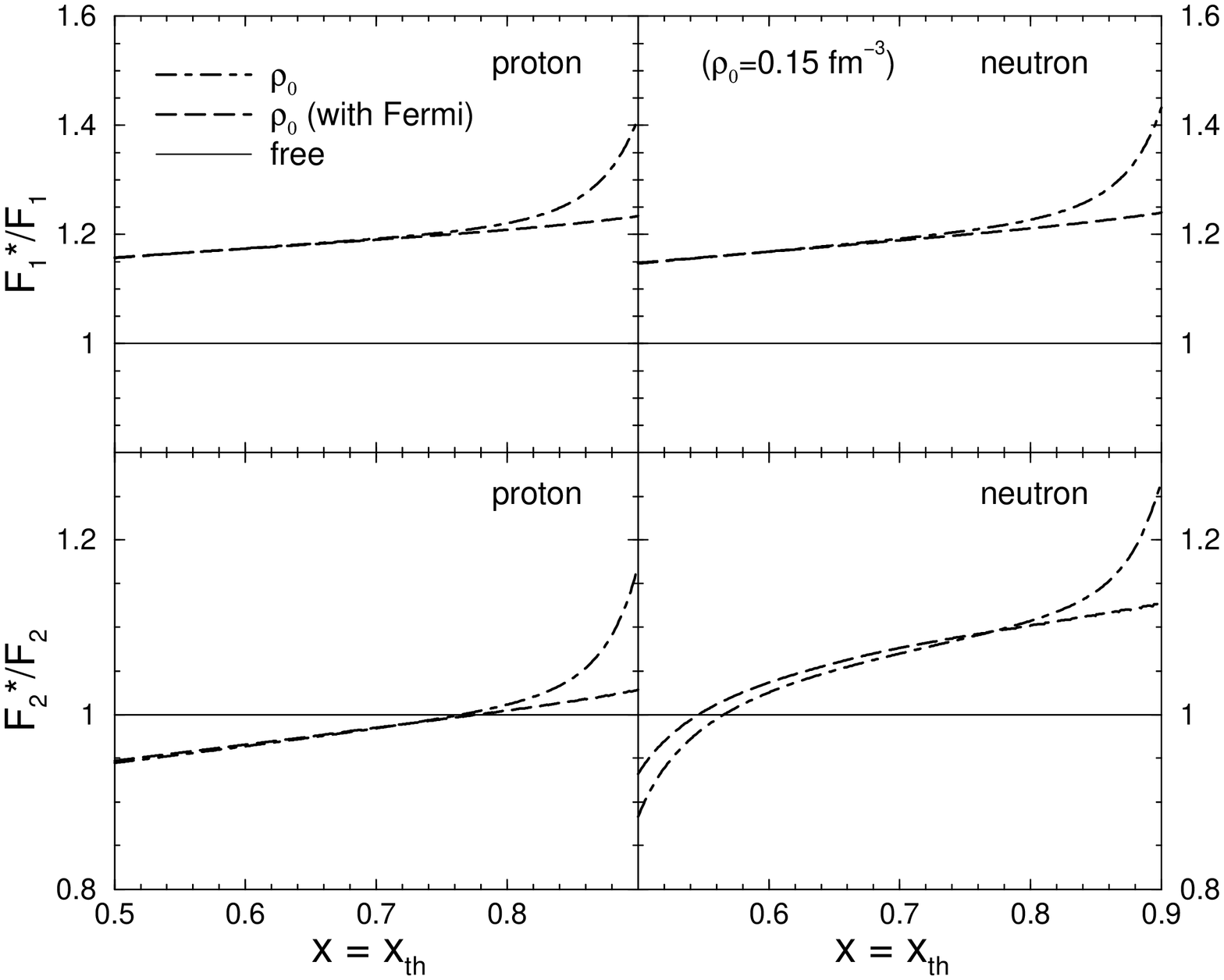,angle=-90,width=16cm}
\caption{Ratios for the charged-lepton scattering structure functions 
$F^*_{1,2}/F_{1,2}$, those extracted using the 
bound and free nucleon form factors. 
Effect of Fermi motion   
is included by the convolution with the 
nucleon momentum distribution~\protect\cite{Miller} for {\it both} the 
structure functions extracted using the free and bound nucleon form 
factors, and then ratios are calculated
(the dashed lines denoted by "with Fermi").
}
\label{fig_F12}
\end{center}
\end{figure}

\begin{figure}
\begin{center}
\vspace*{1cm}
\epsfig{file=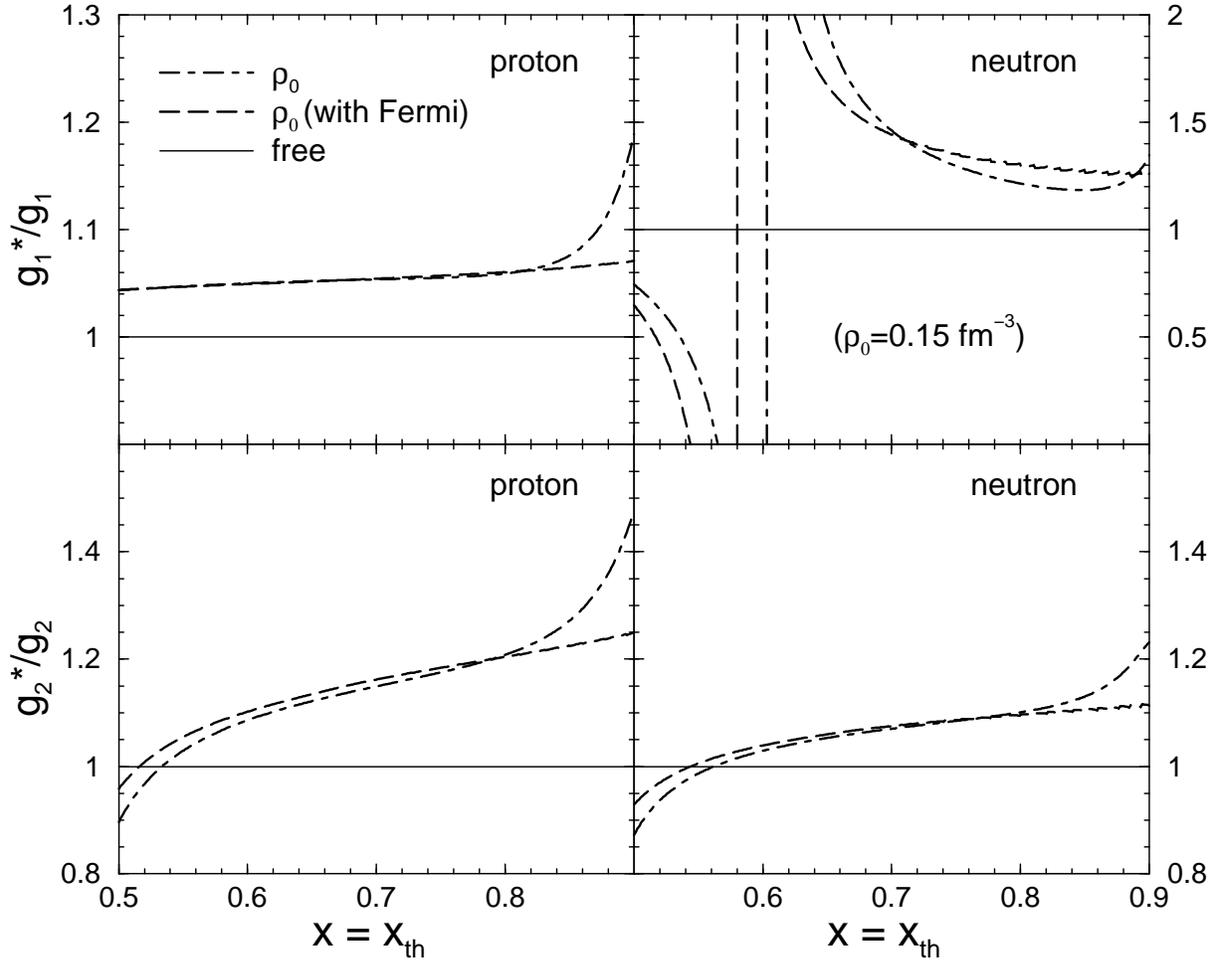,angle=-90,width=16cm}
\caption{Same as Fig.~\protect\ref{fig_F12}, but for $g^*_{1,2}/g_{1,2}$.
}
\label{fig_g12}
\end{center}
\end{figure}

First, we show  
in Figs.~\ref{fig_F12} and~\ref{fig_g12} ratios of
the bound to free nucleon structure functions without the effect 
of Fermi motion (dash-doted lines), calculated for the
charged-lepton scattering for baryon densities $\rho_B =
\rho_0$ with $\rho_0 = 0.15$ fm$^{-3}$. Because $x_{th} =
(0.7,0.9)$ in free space (see also Eq.~(\ref{x_th_medium}))
correspond to $Q^2 \simeq (0.65,2.5)$ GeV$^2$, we regard the
results shown in the region, $0.7 \alt x \alt 0.9$, as the present
local duality predictions. The corresponding $Q^2$ range is also
more or less within the reliability of the bound nucleon form
factors calculated~\cite{He4,QMC_EMff,neuA}. 
In the region, $0.8
\alt x \alt 0.9$, all the bound nucleon structure functions
calculated, $F^*_{1,2}$ and $g*_{1,2}$, are enhanced relative to
those in a free nucleon.
(We have also checked that the enhancement becomes larger as the
baryon density increases.) 
However, the depletion observed in the
EMC effect, occurring just before the enhancement as $x$
increases, is absent for all of them. Probably, the conventional
binding effect, which is not entirely included in the present study, may
produce some depletion~\cite{ST}. (Also recall the conclusion
drawn by Smith and Miller~\cite{Miller}.) Thus, only the effect of
the bound nucleon internal structure change introduced via the
bound nucleon form factors and the pion threshold shift in the
present local duality framework, cannot explain the observed
depletion in the EMC effect for the relevant Bjorken-$x$ 
range $0.7 \alt x \alt 0.85$. 
However, it can explain a part of the enhancement 
at large Bjorken-$x$ ($x \agt 0.85$).

In order to see whether or not the conclusions drawn above are affected 
by Fermi motion, we also calculate the ratios by convoluting   
the nucleon momentum distribution  
obtained in Ref.~\cite{Miller} with the value $\bar{M} = 931$ MeV. 
Namely, we convolute the nucleon momentum distribution with {\it both} the 
structure functions extracted using the free and bound nucleon 
form factors first, and then calculate ratios.
The corresponding results are shown  
in Figs.~\ref{fig_F12} and~\ref{fig_g12} (dashed line, 
denote by "with Fermi").
Note that, because the upper value of 
$x_{th} (x_{th}^{max} \sim 0.91)$ is limited for a reliable extraction 
of the structure functions  
by the reliable $Q^2$ range for the nucleon elastic 
form factors in this study, we had to cut the contribution from the region,  
$x_{th} \ge x_{th}^{max}$, in the convolution integral. 
This would effectively suppress the enhanced part of the 
bound nucleon structure functions.
The obtained results show a similar feature, except the region 
$x_{th} \agt 0.8$. 
However, even the region $x_{th} \agt 0.8$,  
the enhancement feature remains the same. 
Thus, we conclude that, the enhancement  
of the bound nucleon structure functions $F^*_{1,2}$ and $g^*_{1,2}$ in the 
charged lepton scattering obtained, 
is intrinsic and not smeared by the effect of Fermi motion.
This is more obvious in the region $0.7 \alt x_{th} \alt 0.8$. 
In that region, the effect of Fermi motion is nearly  
canceled out in the ratios, as one could expect.

\begin{figure}
\begin{center}
\vspace*{1cm}
\epsfig{file=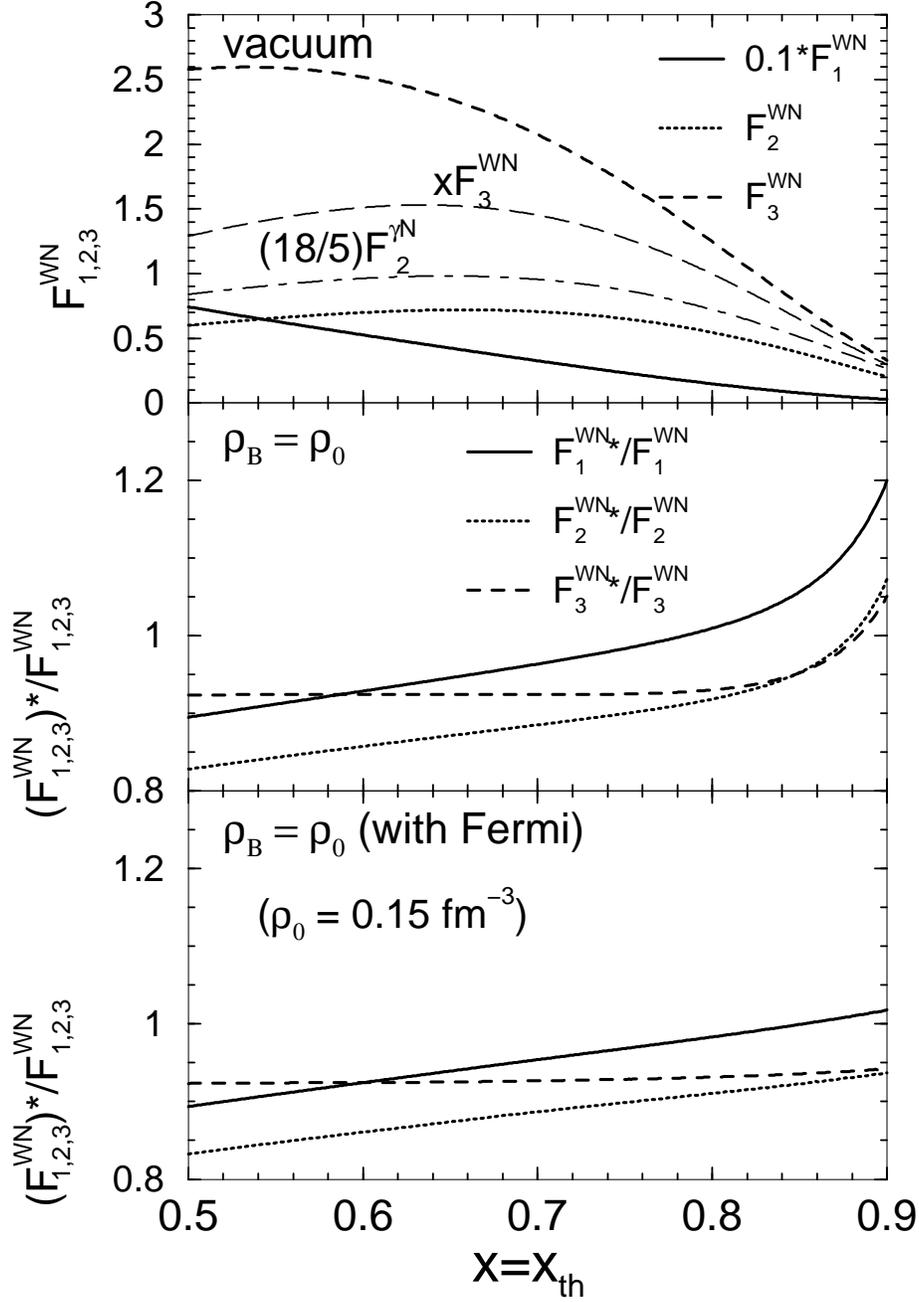,angle=-90,width=12cm}
\caption{Structure functions calculated for (anti)neutrino
scattering (for an isoscalar nucleon), for baryon density 
$\rho_B = 0$ and $\rho_0$. For a reference,
$\frac{18}{5}F_2^{\gamma N} \equiv \frac{18}{5}\frac{1}{2}(F_2^p+F_2^n)$
for the charged-lepton scattering in vacuum is also shown (top panel).
Ratios are shown for both with and without the effect of Fermi motion.  
(See also caption of Fig.~\protect\ref{fig_F12} for 
the effect of Fermi motion.)
}
\label{fig_Fjwmed}
\end{center}
\end{figure}

Next, we show in Fig.~\ref{fig_Fjwmed} the bound nucleon structure
functions calculated from a charged current, $F^{WN*}_{1,2,3}$,
together with those in vacuum for the (anti)neutrino scattering.
For a reference, we show also $\frac{18}{5}F_2^{\gamma N} \equiv
\frac{18}{5} \frac{1}{2}\left(F^p_2+F^n_2\right)$ in vacuum for
the charged-lepton scattering. 
The effect of Fermi motion is included in the same way as that 
was included in the charged-lepton scattering case.
Similarly to the charged-lepton
scattering case, $F^{WN*}_{1,2,3}$ in symmetric nuclear matter are
enhanced at large $x$ without the effect of Fermi motion, 
but only in the region, $0.85 \alt x \alt
0.9$. This is due to the contribution from the in-medium axial
vector form factor $G^*_A$. Although $G^*_A$ falls off faster than
the free space $G_A$ in the range $Q^2 \alt 1$ GeV$^2$, the $Q^2$ dependence
turns out to be slightly enhanced in the range $Q^2 \agt 1$
GeV$^2$, due to the Lorentz contraction of the internal quark wave
function of the bound nucleon~\cite{neuA}. Then, the contribution
from the $Q^2 \agt 1$ GeV$^2$ region gives a suppression. (See
Eqs.~(\ref{F_1WN_x_th}) -~(\ref{F_3WN_x_th}), but neglecting small
contributions from the non-derivative terms with respect to $Q^2$,
which are suppressed by $\sim 1/\tau^2$ as $Q^2$ increases.)
With the effect of Fermi motion, the enhancement and quenching features 
at $x_{th} \agt 0.8$ are less pronounced because of 
the convolution procedure 
applied in the present treatment.

\begin{figure}
\begin{center}
\vspace*{1cm}
\epsfig{file=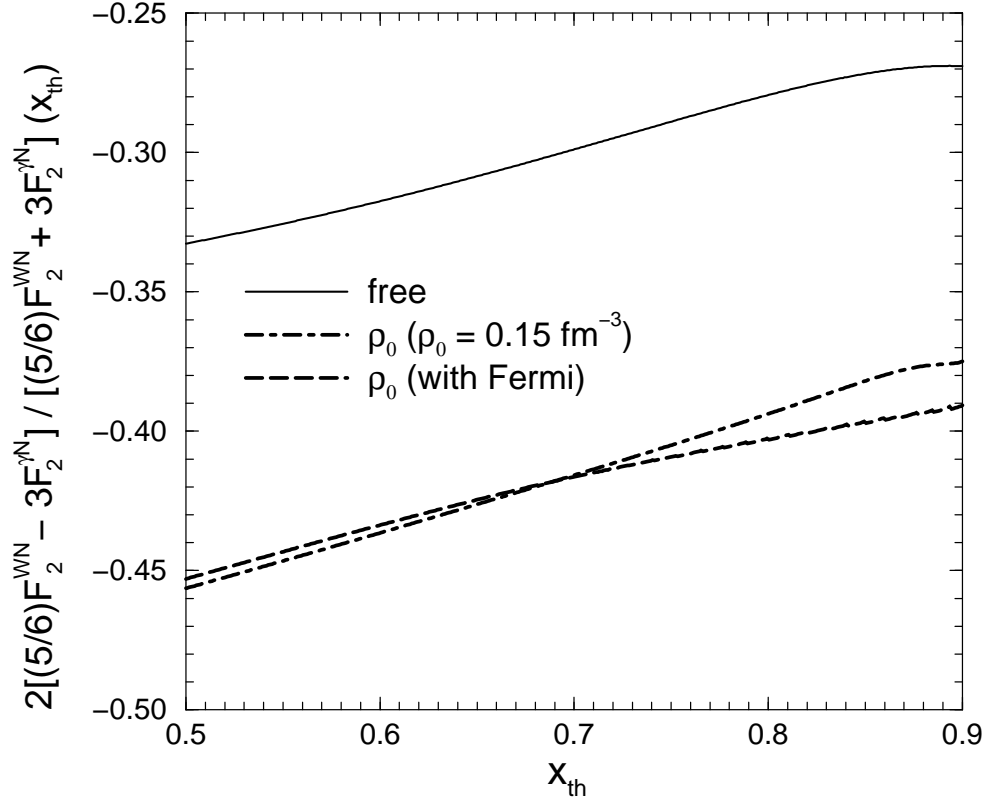,angle=-90,width=12cm}
\caption{Normalized ratios, divided by
$\frac{1}{2}\left[\frac{5}{6}F_2^{WN*}+3F_2^{\gamma N*}\right]$,
for charge symmetry breaking in parton
distributions, for baryon densities $\rho_B = 0$ and
$\rho_0$, with and without the effect of Fermi motion.
(See also caption of Fig.~\protect\ref{fig_F12} for the effect of 
Fermi motion.)
}
\label{fig_F2CSBmed}
\end{center}
\end{figure}

After having calculated $F_2^*$ for both the charged-lepton and
(anti)neutrino scattering, we can now study charge symmetry
breaking in parton distributions focusing on the effect of the
bound nucleon internal structure change, with and without the effect 
of Fermi motion. In free space, it was
studied in Ref.~\cite{FK} based on the local duality. A measure of
charge symmetry breaking in parton distributions at $x = x^*_{th}$
may be given by~\cite{FK}:
\begin{eqnarray}
\left[\frac{5}{6}F_2^{WN*}(x^*_{th})\right.
&-& 3\left.F_2^{\gamma N*}(x^*_{th})\right]
\nonumber\\
&=&
3\left\{\frac{13}{18}\beta^*\left[\frac{d(G^{p*}_M)^2}{dQ^2}
+ \frac{d(G^{n*}_M)^2}{dQ^2}\right]
+ \frac{5}{9}\beta^*\frac{d(G^{p*}_MG^{n*}_M)}{dQ^2}
- \frac{5}{18}\beta^*\frac{d(G^*_A)^2}{dQ^2}\right\}.
\label{CSB}
\end{eqnarray}
In Fig.~\ref{fig_F2CSBmed} we show normalized ratios, divided by
$\frac{1}{2}\left[\frac{5}{6}F_2^{WN*}+3F_2^{\gamma N*}\right]$,
for baryon densities $\rho_B = 0$ and $\rho_0$ with and without the 
effect of Fermi motion. The
results show that the charge symmetry breaking in symmetric
nuclear matter is enhanced due to the effect of the bound nucleon
internal structure change.  
(We have checked that the breaking becomes larger 
as the baryon density increases.)
Note that, because the quantity is the ratio 
by definition, it is very insensitive to the effect of Fermi motion 
in entire region of $x_{th}$ considered. 
Thus, charge symmetry breaking looks to
be more appreciable in nuclei than in the case of a free nucleon,
and/or could affect fragmentation in heavy ion collisions. In
particular, the results imply that the NuTeV anomaly~\cite{NuTeV},
which was observed in the measurements using iron target, would be
enhanced even more than the analysis made~\cite{FK} in free space.
(See e.g., Ref.~\cite{Londergan} for detailed discussions.)
However, the present status of experimental accuracies would not
allow to detect the effect distinctly.

We summarize the results and conclusions of the present study
based on the local quark-hadron duality:
\begin{enumerate}
\item The effect of the bound nucleon internal structure change in
the nuclear medium is to enhance the bound nucleon structure
functions at large Bjorken-$x$ ($x \agt 0.85$) for the
charged-lepton scattering and especially $F_1^{WN}$ in 
(anti)neutrino scattering.
\item The $x$ dependence of the bound nucleon structure functions
obtained for the charged-lepton scattering and $F_{2,3}^{WN}$ in 
(anti)neutrino scattering
is different. Namely, the former [latter] is enhanced [quenched] in the
region $0.8 \alt x \alt 0.9$ [$0.7 \alt x \alt 0.85$].
\item Only the effect of the bound nucleon internal structure change cannot
explain the depletion observed in the EMC effect (for the charged-lepton
scattering) for the relevant Bjorken-$x$ range in this study, 
$0.7 \alt x \alt 0.85$, but it can explain a part of the enhancement
occurring in the larger region of $x$.
\item Charge symmetry breaking in parton distributions in nuclei,
or higher baryon densities, would be enhanced relative to that in free space
due to the internal structure change of a bound nucleon.
\item The conclusions obtained above are insensitive to the effect of 
Fermi motion in the treatment of the present study.

\end{enumerate}
Finally, we again note that the present study have not included 
in a rigorous manner the
conventional nuclear effects, such as binding and Fermi motion.
(This will be investigated in future work~\cite{Saito}.)
However, even solely from the present results,
we can conclude/suggest that,
not only the conventional nuclear effects,
but also the effect of the bound nucleon internal structure change,
may be appreciable in various nuclear structure functions 
at large Bjorken-$x$.

\vspace{1em}
\noindent
{\bf Acknowledgment:}
We would like to thank W. Melnitchouk for the correspondence on
Ref.~\cite{Wally}.
KT was supported by FAPESP (03/06814-8), and FMS was supported
by FAPESP (03/10754-0), CNPq (308032/2003-0) and Mackpesquisa.
%

%
\newpage
\newpage
\newpage
\newpage
\end{document}